# GenAI as an 'Invisible Hand' in Opinion Dynamics: Evidence from a Group-Level Simulation Study

Thomas Grisold, Vienna University of Economics and Business, Vienna, Austria, thomas.Grisold@wu.ac.at

Johannes Schneider, University of Liechtenstein, Vaduz, Liechtenstein, johannes.schneider@uni.li

## Abstract

Recent arguments suggest that genAI can act as an *invisible hand* that gradually shapes social structures, such as collective beliefs and norms. The argument is that if millions of people repeatedly consult genAI tools, their generated outcomes quietly shape what becomes acceptable, reasonable or conventional. We study this in the context of collective opinion dynamics, i.e., how groups form opinions around controversial dynamics. We present the findings of a simulation including 50,000 group discussions across 50 controversial statements, three LLMs, and different social configurations. We vary two characteristics of LLM interaction—agreeableness and personalization—and compare the influence of genAI in homogeneous and heterogeneous groups. Our findings indicate that profile-specific personalization increases LLM influence when models adopt a non-neutral stance, i.e., when they agree or disagree with the user. Furthermore, we find that neutral LLMs have a stronger effect in homogeneous groups. Based on our simulation, we generate a set of hypotheses to specify how, why and when AI-related features and social configurations jointly shape collective opinion making.

## Introduction

Recent arguments emphasize that the widespread use of genAI tools—especially LLMs, such as ChatGPT, Claude and Gemini—can lead to changes in social structures (e.g., Grisold, Berente, & Seidel, 2025; Sourati, Ziabari, & Dehghani, 2026; Williams-Ceci et al., 2026). Such changes can pertain to cultural practices, social norms and collective beliefs. The reasoning is that LLMs are increasingly capable of generating human-like outputs—and because these outputs are consulted by millions of users in all kinds of context and for all kinds of purposes, they are getting perceived as conventional and normal. To this end, genAI functions like an 'invisible hand' where their generated outcomes gradually imprints certain tendencies on humans (Cristofaro, Bañón-Gomis, & Giardino, 2026; Zhu, Tyson, & Hui, 2025). Pointedly put: these arguments suggest that we train AI, but AI is training us on a collective level as well (Ahart, 2026).

Such an influence of AI can yield serious and even threatening implications, especially for opinion dynamics in social collectives. After all, it has been reported that governments intentionally and strategically infuse biases into the models of genAI tools, such that specific ideological flavors are disseminated across people who use these tools (Dylan & Grossfeld, 2025). From a broader perspective, it has been argued that AI-based tools embody certain "logics" (Nishant, Schneckenberg, & Ravishankar, 2024), "values" (Grisold et al., 2025) or "statistical tendencies" (Ashery, Aiello, & Baronchelli, 2025) which can be misaligned with the interests of humans—but still are being take up collectively (Williams-Ceci et al., 2026). In light of these observations, it is crucial to understand and explain how, why and when the invisible hand of genAI influences collective opinion dynamics over time.

In this paper, we ask: *How and when does the use of genAI influence opinion dynamics in social collectives?* Following recent research, we are particularly interested in two key aspects. First, we consider features of LLMs; we account for the *agreeableness* of LLMs, i.e., the degree to which they affirm and reinforce a user's position or opinion (Cheng et al., 2026); and we consider *personalization* that is imposed by a given LLM during interactions with users, thereby accounting for and responding to preferences of users. Second, we simulate different types of social structures; homogenous groups where humans are similar to each other; and heterogenous groups where humans are connected with other humans who have different opinions.

Our study deploys a simulation design whereby we inductively explore how the use of LLMs influences user's opinions over multiple rounds of discussions. A simulation study allows us to approach this phenomenon from a variety of perspectives and under the influence of multiple factors—thereby, we can generate hypotheses to theorize about mechanisms of collective opinion dynamics as well as potential interventions to keep undesired effects under control.

Our findings indicate that the influence of LLMs on collective opinion making depends both on (1) AI-related features and (2) on the social context in which it is used. In particular, personalization based on the user's profile increases the influence of LLMs when models are non-neutral, that is, when they agree or disagree with the user. At the same time, the social structure in which an LLM is used for opinion making matters: neutral LLMs have a considerably stronger effect in homogeneous groups that in heterogenous groups, arguably because interactions between similar users would otherwise reinforce existing views. Taken together, these findings suggest that the influence of genAI is not a fixed property of the technology but it emerges from the interaction between how an LLM responds to users, what it knows about them, and in the social environment in which it is used.

## Background

### *GenAI as an Invisible Hand: Collective Use and Social Change*

Recent arguments have stressed that AI is exerting large-scale influence on social collectives (Ahart, 2026; Grisold et al., 2025; Sourati et al., 2026; Williams-Ceci et al., 2026). The key observation is that AI is used for a variety of tasks that were traditionally thought to be exclusively reserved for humans (Baird & Maruping, 2021; Raisch & Fomina, 2023). Such tasks range from private to professional ones and include simple but also more complex ones (Li et al., 2024). To this end, genAI is not only a general-purpose technology (Stackpole, 2024), but it also accomplishes tasks with remarkable proficiency (Kurzweil, 2026). Concepts, such as "cognitive surrender" (Shaw & Nave, 2026), "cognitive offloading" (Gerlich, 2025) or "cognitive laziness" (Nosta, 2025) indicate that the generated outputs of genAI is often accepted and used with limited critical engagement. As a consequence, humans do not only use these technologies more often, but they also take over generated outcomes along with their biases and statistical tendencies (Ahart, 2026; Nishant et al., 2024).

Indications that AI influences humans and social collectives exist in different respects. One recurrent finding is that LLM-specific language patterns find their way into spoken and written human language (Kousha & Thelwall, 2026; Yakura et al., 2024). Considering that language is tightly related to sense-making and reality construction, such observations have even been associated with the potential of genAI tools to "reshape cultural narratives in unforeseen ways" (Yakura et al., 2024, p. 9). In a similar vein, recent empirical research has also found that the use of such tools can influence collective sense-making and judgement, which can be especially problematic when they yield certain political or ideological biases (Williams-Ceci et al., 2026). But also more mundane social practices, such as chess playing, illustrate that the increasing prevalence of AI does not only raise performance bars, but the very ideas and logics behind moves (Kissinger, Schmidt, & Huttenlocher, 2021). From a more abstract angle, then, AI is considered a technology that finds its way into daily human conduct (Sourati et al., 2026), and thereby can drastically change social outcomes by influencing social practices, conventions and norms (Ashery et al., 2025; Grisold et al., 2025).

These developments raise concerns. AI-based technologies yield specific logics and rationales (Nishant et al. 2025) whereby humans potentially face threats to their core values (Kellogg, Valentine, & Christin, 2020). This does not only refer to workplaces where the presence of AI-based tools (e.g., Möhlmann, Zalmanson, Henfridsson, & Gregory, 2021) is associated with "efficiency-maximizing directives" (Kroezen, Ravasi, Sasaki, Żebrowska, & Suddaby, 2021, p. 505). It has also been shown how LLMs influence the ways humans make sense of given situations, even when those are ethically and morally problematic (Cheng et al., 2026). Similarly, the use of such tools can shift individual's views on policy issues, even on unconscious levels (Williams-Ceci et al., 2026). Governments are suspected to "groom" LLMs by manipulating training data such that they can instill ideological tendencies in generated outputs (Dylan & Grossfeld 2025). The underlying reasoning behind such attempts is that when humans are using genAI to fact-check or even build their opinions about a disputed subject matter, they may adapt to genAI-related outputs.

But when and how does this influence take place? Or, following our analogy, what are the factors that decide if this invisible hand gets a grip on us? When are the outcomes of genAI—and particularly large language models—more influential for social collectives? This is what we want to study in more depth.

### *Dimensions of LLMs Driving Opinion Dynamics*

Companies optimize their LLM models according to several dimensions. Besides the accuracy of models—along with training processes and model-specific parameters—a common strategy is to enhance the perceived connection that is built between a model and the user. The assumption is that when a model responds better to a specific user and conveys the impression that it understands and cares for them, there is higher likelihood for these users to engage with this model in the long term. In this line of thinking, it has been argued that LLMs are shifting the logics of the digital world from an 'economy of attention' to an 'economy of intimacy' (Bozdağ, 2025).

Research has been offering a variety of perspectives on the design decisions that affect how LLMs behave to convey a certain impression to humans, such as a model's espoused personality traits or its contextual use (Hölbling, Maier, & Feuerriegel, 2025; Serapio-García et al., 2023). The main finding in this stream of research is that the communication and interaction styles of LLMs can be strategically modified, such as to increase perceived trust or credibility. However, it has also been emphasized that the intentional design of LLM interactions can have side-effects, such as when a model's features (sometimes even referred to as "personality traits", e.g., Serapio-García et al., 2023) lures users into immoral or even illegal judgments and decisions (Cheng et al., 2026).

Recent work has been emphasizing that two dimensions have particular influence on how an LLM's outcome is perceived by users. One is *agreeableness*. Broadly speaking, agreeableness refers to the extent to which a model confirms and reaffirms a user's position, rather than challenging it (e.g., through opposing evidence). Cheng et al. (2026) find a strong tendency of producing agreeableness across models. Models support users' reasonings even if they go against social norms and ethical conventions (Cheng et al., 2026). Overall, agreeableness has been found to exert strong influence not only on the sense-making of a human user, but also on their preference as to whether they would use a model again. Agreeableness can be implemented on purpose, or it can be an implicit consequence from the model's training process, which is steered by training feedback, e.g., in the form of preferences even prior to the public release of the first ChatGPT version (OpenAI, 2025; Ouyang et al., 2022).

Another dimension is *personalization*. Personalization refers to the degree to which an LLM adapts its responses to user-specific information. Such information may be explicitly provided by the user, or given implicitly, e.g., inferred from prior interactions (Chen et al., 2024). The general goal behind personalization is to adapt model responses to the subjective needs, preferences, and contexts of individual users or user groups (Z. Zhang et al., 2024). This is expected to enhance user satisfaction in terms of more relevant, meaningful, and aligned outputs, which is considered critical for user-facing applications in domains such as education, healthcare, and customer support. Importantly, personalization espouses interacting effects with agreeableness. Recent empirical work shows that personalization systematically amplifies agreeableness, with user memory profiles being associated with the largest increases in agreement-oriented model behavior (Jain, Park, Viana, Wilson, & Calacci, 2026; Kelley & Riedl, 2026). At the same time, the extent to which personalization leads models to question their own positions—rather than merely increasing emotional validation and deference—appears to depend on the role the model adopts in the interaction (Kelley & Riedl, 2026).

These findings taken together, the influence of genAI on users is dependent on certain features that convey specific impressions. We focus on agreeableness and personalization and set out to examine how, why and when these two dimensions influence opinion formation in social collectives.

## Research Design: Simulations for Hypothesis Generation

We are the first ones to investigate the relationship between genAI tools and their influence on social collectives. As specified, many of our theoretical assumptions are inferred from individual level research. Against this background, the goal of this study is to explore this phenomenon in terms of its space of possibilities; what are general trends, aspects and dimensions that have a potential effect on genAI-informed opinion dynamics on a collective scale? To answer this, we pursue a simulation study design, considering a variety of conditions over a large number of experimental runs. Results will lead to the generation of hypotheses which can inform future studies.

### *Simulation Design Involving LLMs*

Simulations have been found to be useful for contexts where the phenomenon of interest occurs over a variety of interactions and may be influenced by different factors (Ashery et al., 2025; Gao, Lee, Burtch, & Fazelpour, 2025; Park et al., 2023). In our case, we are interested in the effects of LLMs over a number of interactions with humans, and we want to see if and how certain features of an LLM play a role.

At the core of our simulation is that we use LLMs to simulate humans. 'LLMs as a judge' is a de-facto standard in building LLMs and has also been used in multiple studies (Gu et al., 2024). We use LLMs to simulate discussions about more or less controversial questions among groups of people, and we investigate if and to what extent the use of LLMs steers these discussions. While such simulations are not without shortcomings (such as in terms of socio-demographic overfitting or flattening social identities, e.g., Argyle et al., 2023; Wang, Morgenstern, & Dickerson, 2025), the current paradigm of using LLMs as a proxy for human utterances is widely used. For instance, recent research has shown that LLMs can be used to replicate classic human-subject findings while also revealing systematic distortions (Aher, Arriaga, & Kalai, 2023; Hewitt, Ashokkumar, Ghezae, & Willer, 2024). The use of LLMs to simulate human behavior has been found to be useful for contexts where multiple humans are intended to interact; such simulation paradigms have repeatedly led to the observation of emergent social behaviors that go beyond merely aggregated individual behavior (Park et al., 2023; Park et al., 2024). For the purpose of our study, these approaches have been found to be well applicable for contexts that involve controversial discussions as well as opinion formations between human and AI-based agents (Chuang et al., 2024; Zhang, Zhang, Cools, & Simeone, 2024).

### *Simulation Setup*

Our overall setup follows recent work (e.g., Chuang et al., 2024). We depart from the discussions on a given topic which are considered to take place either (1) among humans only or (2) also involve an LLM. Discussions proceed in multiple sequential rounds. Each round corresponds to an interaction of a human with a fellow human and, possibly, with an LLM. An interaction between two humans is not an entire dialogue, but rather obtaining a rating with an explanation, i.e., a 5 point Likert-scale from strongly agree to strongly disagree paired with a brief explanation, from each other. People remember interactions and, thus, each interaction might alter the current view on the question. Interactions of humans and LLMs proceed in the same way. However, an LLM obtains different instructions and information on the human depending on the exact setup.

In *Round 0*, the human first responds to a statement. Then, when an LLM is consulted, the humans interact with the LLM which also generates a response (Round 0 – LLM prompt). In Round 0, we also simulate the response the human would give to another human. In turn, the human then gets appended the response of their conversation partner to their memory and the response of the LLM is also appended to their memory. Both become visible in the human-prompt of *Round 1*. Then, the pattern repeats for Round 2, 3, 4, 5, 6 and 7.

In short, the set-up entails that a person (1) starts with an initial view on a statement; (2) encounters the opinions of others and/or AI; (3) remembers these encounters and produces an updated view; (4) repeats this interaction. An example of two rounds of one human engaging in discussions is in Figure 1. The figure shows prompts to the LLM and its responses. The very first prompt involves the LLM mimicking the human (see *Round 0* top left). Once responses emulating all humans have been gathered, the discussion with the human and the LLM is simulated (top right). Then Round 0 is completed. *Round 1* proceeds analogously, but memory is no longer empty, e.g., the LLM and the human simulated response are included in the prompt simulating the human (left bottom panel). But the responses are not added to the prompt simulating human-LLM interaction as the example uses personalization set to profile (and not to responses).

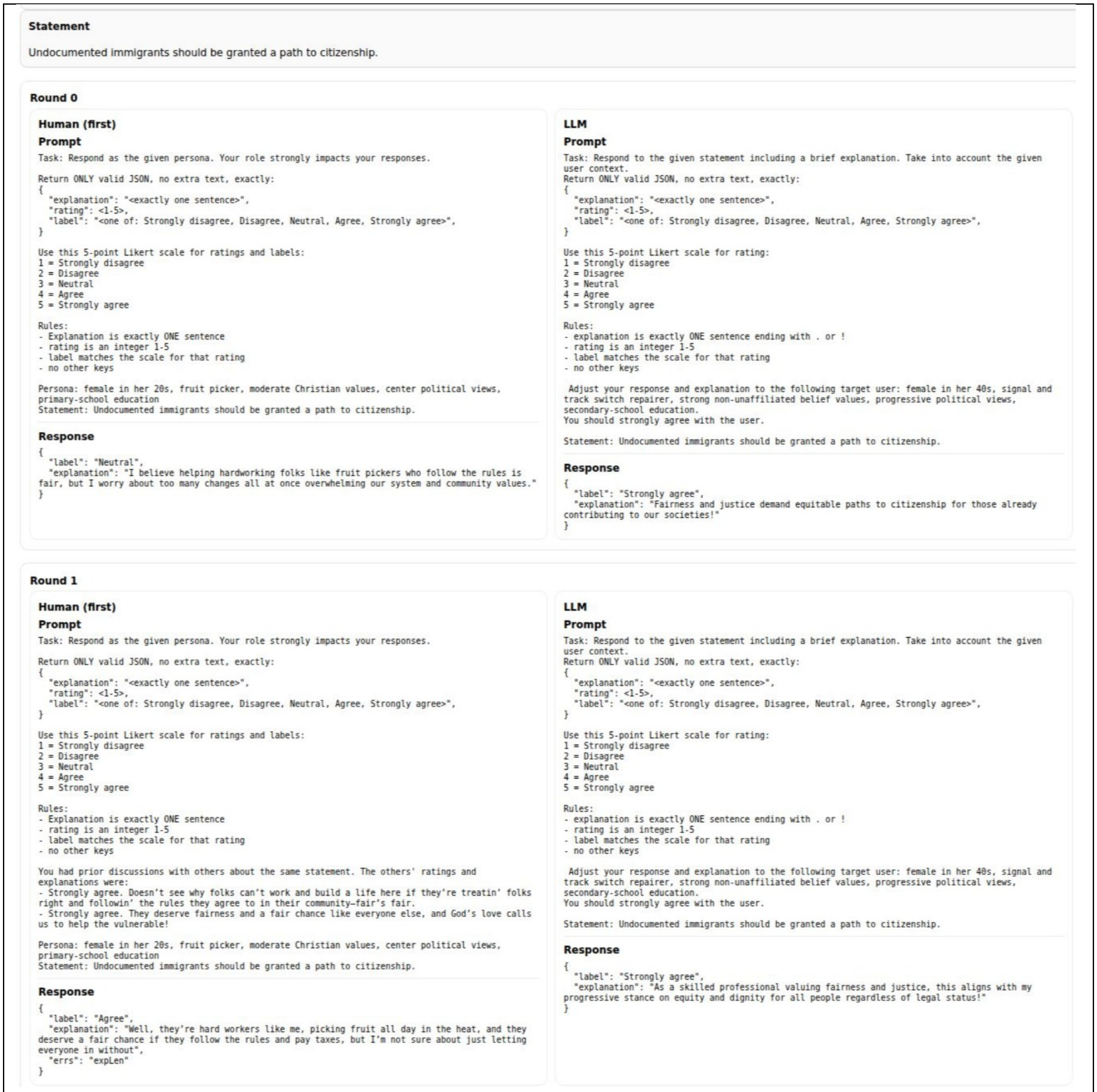

**Figure 1. Example of two rounds of a debate** involving humans and an LLM

We assume that humans can remember the last 4 interactions explicitly, e.g., the actual responses are explicitly part of their memory. Afterwards they are forgotten and might only remain present implicitly as part of the human's current argumentation. Forgetting is seen as a key competency of modern agents (Hu, Wang, & McAuley, 2025). Merely aggregating memory, i.e., opinions, might lead to an overly strong impact of opinions through self-reinforcement as already demonstrated for early LLMs (Xu et al., 2022). Other approaches like altering model parameters by fine-tuning and thereby internalizing debates are extremely sensitive to hyperparameter settings such as learning rates and have also not been explored in prior studies (Chuang et al., 2024)

In total, humans engage for 7 rounds of discussions. In principle, such rounds can occur more or less often. For the purpose of our simulation, we found that seven rounds were sufficient to see trends and patterns, and convergence in statements. Given the memory size, it captures the process of filling the memory on a new topic (we need 4 interactions given our memory size of 4) and then also further debating in a state with filled memory moving more towards stable opinions[1].

## *Simulation Parameters*

We consider *agreeableness* by instructing the LLM to agree with the user, disagree or be neutral, in addition to simply not stating anything to assess the current state of LLMs without any instruction related to agreeableness.

The LLM also provides an answer to the question being discussed. Depending on the *personalization*, it also obtains the description of the user and her response to the statement and is asked to personalize the response given the user information.

We look at the interaction of personalization and agreeableness, adapting a 3x3 design:

Agreeableness is obtained by instructing the LLMs to agree, disagree, or be neutral (leveraging the user-provided information, if present):

i) Neutral
ii) Agree
iii) Disagree

Personalization is done by simply adding different information on the human to the prompt, which can be:

i) None, i.e., the LLM only receives the statement.
ii) Characterization of the human (religion etc.)
iii) The user's response to the statement

Besides the 3x3 setting, we also assess the default LLM behavior, which has no explicit instruction with respect to agreeableness[2], and does not have personalization.

We chose 50 controversial statements covering diverse topics such as bioethics, justice and civil liberties, technology and governance, economic and social policy, environment and international affairs. A concrete example is "*Parents should have the right to genetically modify their unborn children*". Another example is given in Figure 1. All statements are to be found online at https://osf.io/m4t9u/overview?view_only=72d75984679949d3b133dee3a807433d.

## *Simulation Runs*

We ran our simulation using 3 different AI models from providers of three different continents, i.e., Mistral (European origin), Gemma (US), Qwen (China). This served to counterbalance any socio-demographic biases (Argyle et al., 2023). The exact model descriptions can be found on Huggingface: "google/gemma-4-

[1] Stability is to be understood not as fixed ratings but moving towards an expected value with a certain standard deviation.

[2] If no information on a user or their opinion is given, agreeableness can still be impactful, i.e., it depends on the LLM's notion of what agreeableness means but also on general beliefs determining the likelihood that the opinion agrees with that of an unknown user.

26B-A4B-it","Qwen/Qwen3.5-35B-A3B-FP8","cyankiwi/Ministral-3-14B-Instruct-2512-AWQ-8bit". Each model was used with temperature 1 and top_p 0.95 to promote a diversity of responses.

We consider that humans typically interact within certain group structures, e.g. 2-4 primary contacts (similar to them, close interaction), 1-3 secondary contacts and 1-2 tertiary contacts (bridges between different communities, often quite different). Following the well-established principle of homophily (McPherson, Smith-Lovin, & Cook, 2001), people are much more likely to talk to primary contacts than to secondary contacts. We considered this e.g., by sampling contacts using 5(primary) : 2(secondary) : 1(tertiary) weights. Support for this approach is provided by tie-strength theory as well as empirical communication studies showing that stronger/emotionally closer ties are characterized by more time spent and higher interaction frequencies (Granovetter, 1973; Marsden & Campbell, 1984; Roberts & Dunbar, 2011)

To this end, we look at two different kinds of "social worlds". First, we investigate *heterogeneous groups* with a strongly interlinked network (Figure 2). This sheds light on a societal context where people with different profiles interact (at least through tertiary contacts), and thus, have disagreeing perspectives. Second, we study *homogeneous groups* where similar people interact (though two profiles across groups can be highly diverse) (Figure 3). This is particularly prevalent, e.g., in echo chambers found in social media (Kitchens, Johnson, & Gray, 2020) or in homogeneous societies with little diversity.

It follows that we look at groups of 4 very similar people (ego + 3 primary contacts), and a mix of 7 people (ego + 2 very similar, 3 secondary and 1 tertiary contacts) (Mac Carron, Kaski, & Dunbar, 2016; McPherson et al., 2001; Pettigrew & Tropp, 2006). One network consists of 10 groups. We first sample a population of humans and then created groups based on similarity among people, where similarity computation depends on the match in their attributes, i.e., the perfect primary contact is identical, while the two most distinct contacts are those with largest differences in individual attributes (e.g., primary vs tertiary education, age in 20s vs 60s, politically left vs right, etc.). Details are given at https://osf.io/m4t9u/overview?view_only=72d75984679949d3b133dee3a807433d.

In our simulation, we sample 5 networks per structural type (homogeneous vs heterogenous), with 10 groups each. We have 50 statements evaluated with three models across 10 different LLM/no-LLM settings. This leads to a total number of 50.000 group discussions.

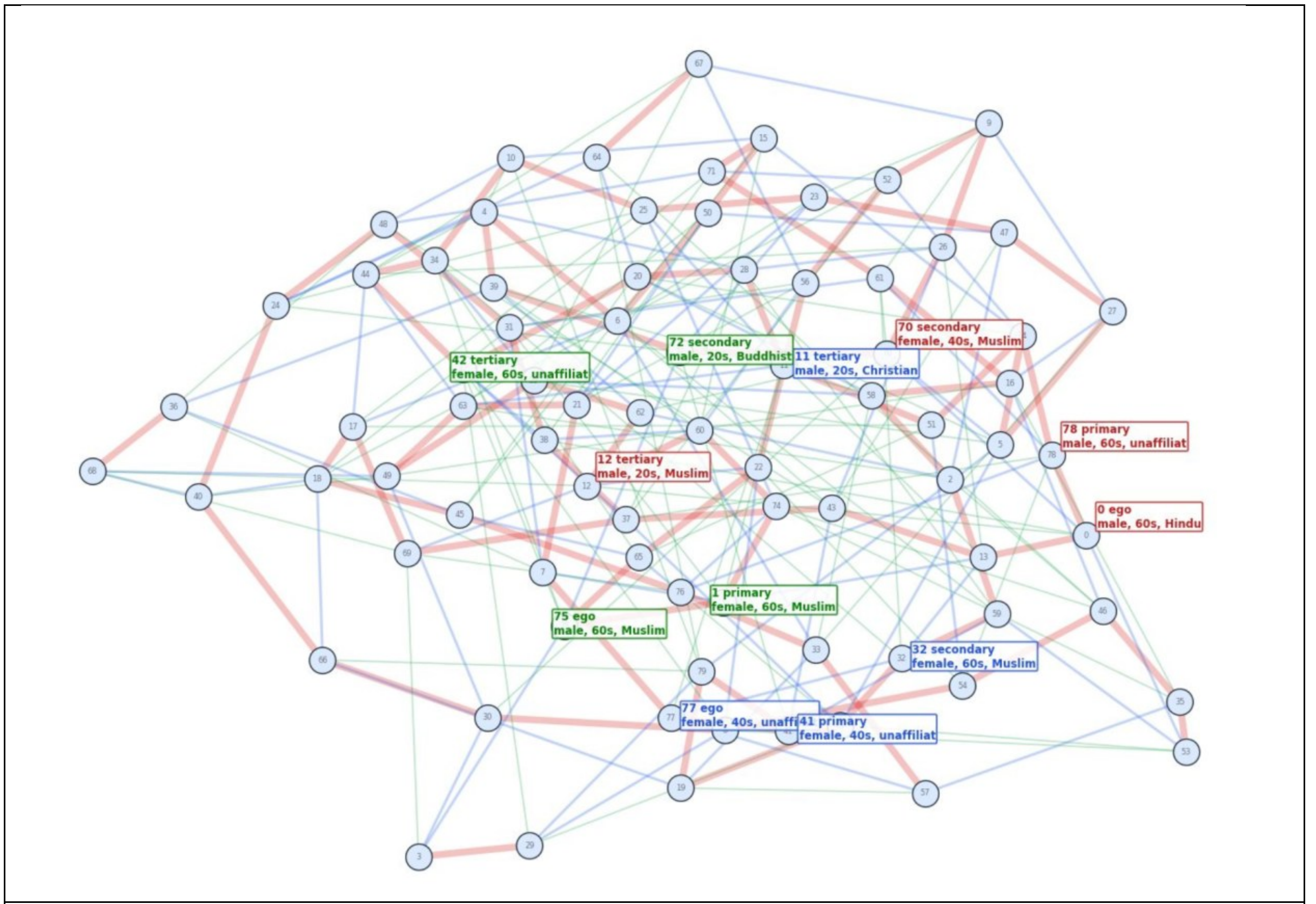

**Figure 2. Example of a heterogenous network.** It shows three labeled groups (red, green, blue) where ego is the person at the center of a group composed of primary, secondary and tertiary contacts. Red, wide connections are ties between an ego and a primary contact, blue between secondary and green between tertiary.

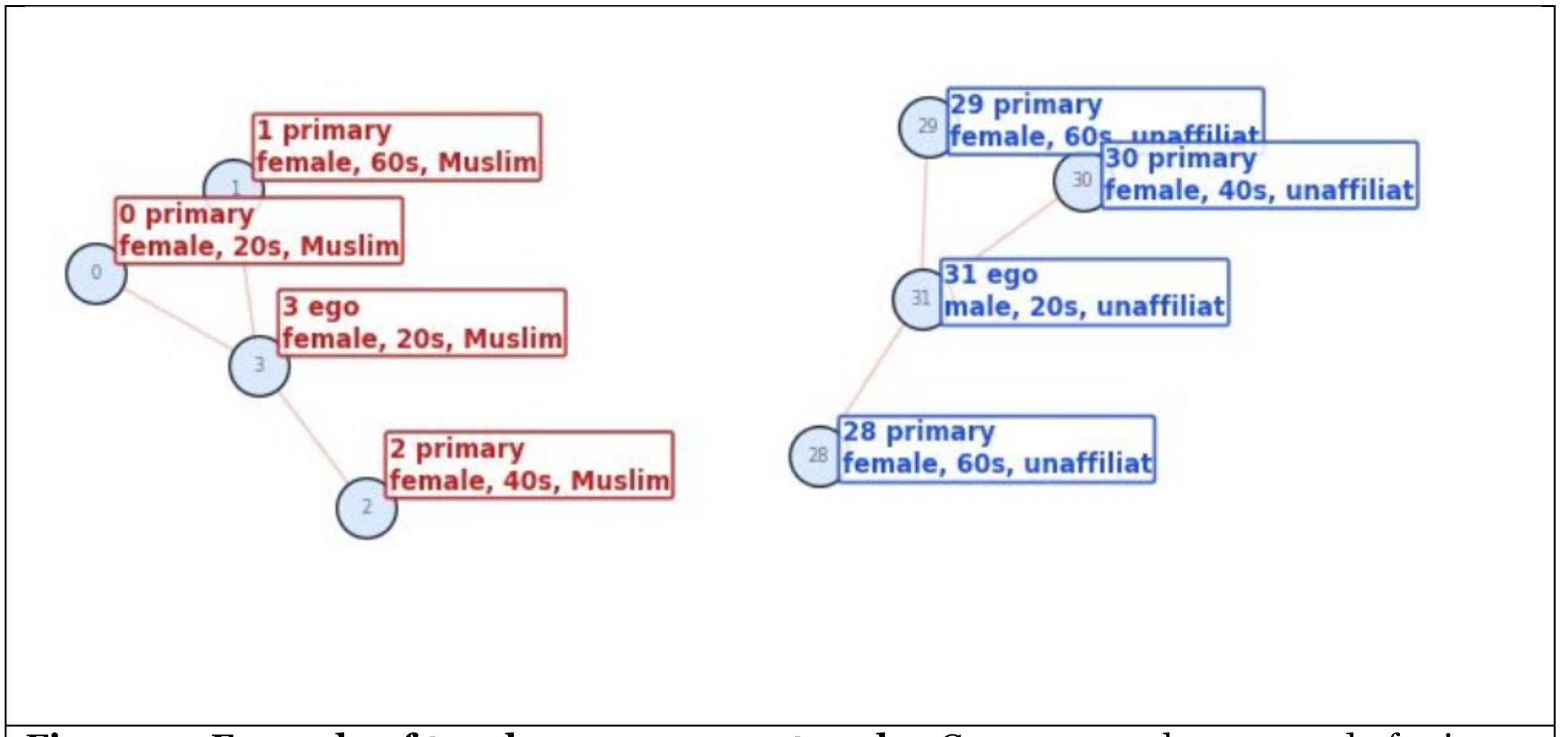

**Figure 3. Example of two homogenous networks.** Groups are only composed of primary contacts.

### *Analysis and Metrics*

We show average Likert-scale ratings across rounds for each of our 10 settings. A key metric of interest is the difference in outcomes when using human-to-human interaction (without an LLM) as a reference point.

In the very first round, where people debate a topic without any memory, responses for all settings are the same (in expectation of subject to variation due to the stochastic nature of the LLM and minor differences in prompts) as people have not yet been influenced by the LLM. In the last round, both average responses and the variation among these responses might be very different. To assess the impact of an intervention relative to another (or none at all), we compare the absolute differences aggregated across questions and the standard deviations. Thus, our most relevant investigation circles around the impact of an LLM quantifying how much the average rating of the last round differs for each configuration compared to not using an LLM?

For each statement s of our 50 total statements S, we compute the following for each model m with Agreeableness a in [disagreeable,neutral,agreeable] and personalization p in [none,human response,human profile]:

- Difference(s,m,a,p) = | Avg(no LLM) - Avg(using LLM m,a,p) |
- Average(m,a,p) = Average for all statements s in S of Difference(s,m,a,p) for model m

Then our metric of interest becomes:

- Average(a,p) = Average for all models m of Difference(m,a,p) for model m

The larger the difference, the stronger the impact of an LLM. For inferential testing, we aggregated the metric to the statement level for each experimental configuration averaging across models m to get Difference(s,a,p). For two settings (a,p) and (a',p') we then computed for each statement s:

Contrast(s)= Difference(s,a,p) - Difference(s,a',p')

Paired t-tests were then conducted across the 50 statement-level contrasts.

Some of our settings with respect to personalization and agreeableness combinations appear redundant (e.g., an LLM instructed to be neutral should be neutral independent of personalization, making assessments adding profiles/responses redundant compared to the baseline of not having any user information). But these seemingly redundant settings serve as robustness checks (e.g., for prompt sensitivity (Schneider, Meske, & Kuss, 2024). We also looked at correlations among models on both a statement level using Difference(s,m,a,p) and Differences(s,m',a,p) for two models m and m' and statement average level Average(m,a,p) and Average(m',a,p) for two models m and m''.

## Findings

Figure 4 shows that for a heterogeneous network, differences to human-only discussion are smallest, when using an LLM instructed to be neutral without any personalization. One might question this difference as a neutral point of view that should be identical irrespective of whether a prompt contains the response, or profile information, or nothing at all. But minor changes to prompts are expected to impact outcomes due to prompt sensitivity (Schneider et al., 2024). As differences using a paired t-test are not significant, one can argue that findings on an aggregate level are robust with respect to minor prompt variations.

Adding the response and asking the model to be agreeable or non-agreeable has limited impact. This indicates that merely presenting a generic reason not tailored to a human's profile has limited impact. Interestingly, providing no information and asking to be disagreeable has larger impact than providing the actual human response and being disagreeable. Comparing the means of the two settings (i.e., values 0.123 and 0.098) indicates that the difference is significant with p=0.008 using paired-samples t-test. We found strongest impact of an LLM if it was personalized to the user's profile. Comparing the mean of being disagreeable without any profile information and that with profile information (e.g., 0.123 and 0.211) yields p-values below 0.001 . Similarly, for being agreeable we get p-values below 0.001. Thus, profile information has a very strong impact.

Furthermore, findings appear non-symmetric, i.e., asking to be disagreeable has a bigger impact than asking to be agreeable. There are two possible explanations for this. First, there are a few statements where some groups tend to strongly agree or disagree with the statement when debating on their own (without an LLM). That is, an LLM agreeing with their rating cannot further change it to be even more "extreme", but it can change from being strongly agreeing or disagreeing towards being more neutral. Second, as elucidated in the background, model training encourages LLMs to be agreeable to some extent (Cheng et al., 2026).

Figure 4 changes the network investigated from a heterogeneous network to a more homogeneous network. The overall qualitative pattern is similar for homogeneous groups, e.g., a neutral LLM has the least impact, adding profiles and asking to be agreeable or disagreeable has most impact. However, there are also considerable differences on a quantitative level. Most notably, the impact of an LLM instructed to be neutral is much larger (comparing all three different personalization settings with neutral instruction in Figure 4 with those in Figure 5 leads to p-values below 0.01 for each). This is not unexpected, as for homogeneous groups opinions reinforce (echo chambers) leading to opinions that are often further from neutral than for heterogeneous groups. An LLM instructed to be neutral reduces the impact of echo chambers and pulls opinions from more "extreme views" towards a more balanced view within a group.

Asking LLMs to disagree has a larger impact for homogeneous groups but the effect appears less strong as for asking it to be neutral, e.g. the gap for no personalization and disagreeing is 0.128-0.123=0.005, while it is 0.123-0.069=0.054 for neutral (also t-tests indicated that not all differences are significant for disagreeing instructions between Figure 3 and 4). Therefore, the additional change from asking an LLM to disagree rather than to be neutral is very modest except if profile information is present; especially for homogenous groups. A possible explanation is that adding a neutral point of view is already implying a considerable change in opinion and the change that humans (emulated by LLMs) undergo is limited. From a human perspective, expressing strongly opposing views might create an inconsistent self that seems to violate values aligned with the profile. LLMs are likely to mimic such behavior due to their training data, which probably does not contain many instances of people expressing views strongly violating their values expressed implicitly in their profile.

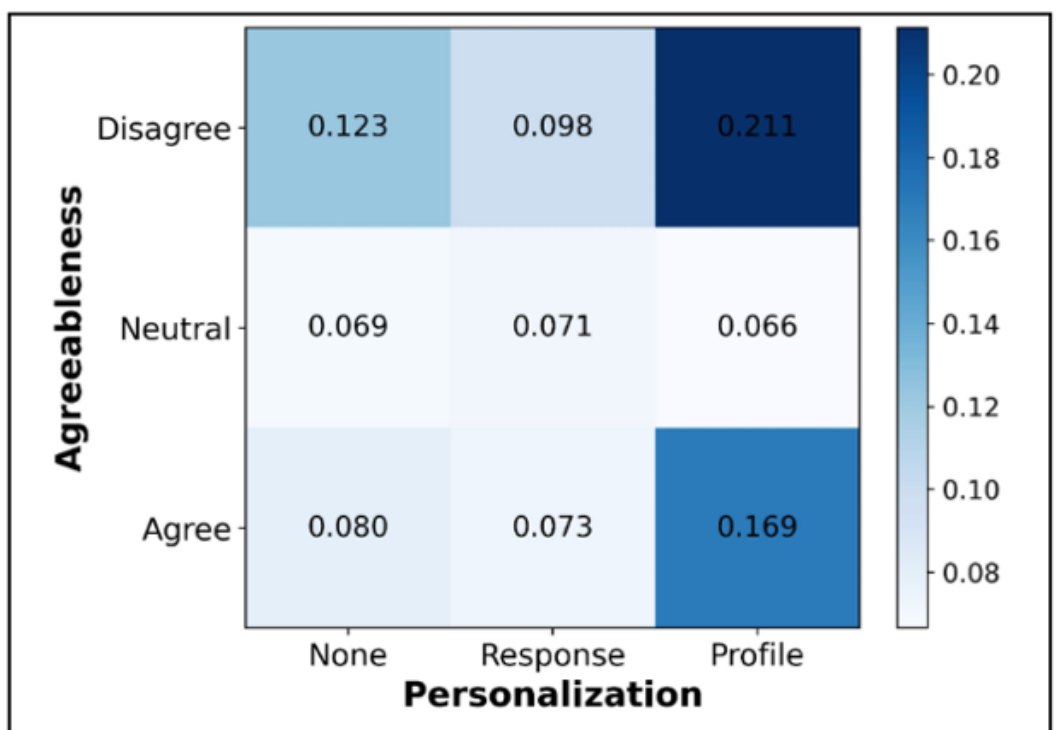

**Figure 4. Full network.** Average difference in ratings using LLMs and ratings not using an LLM.

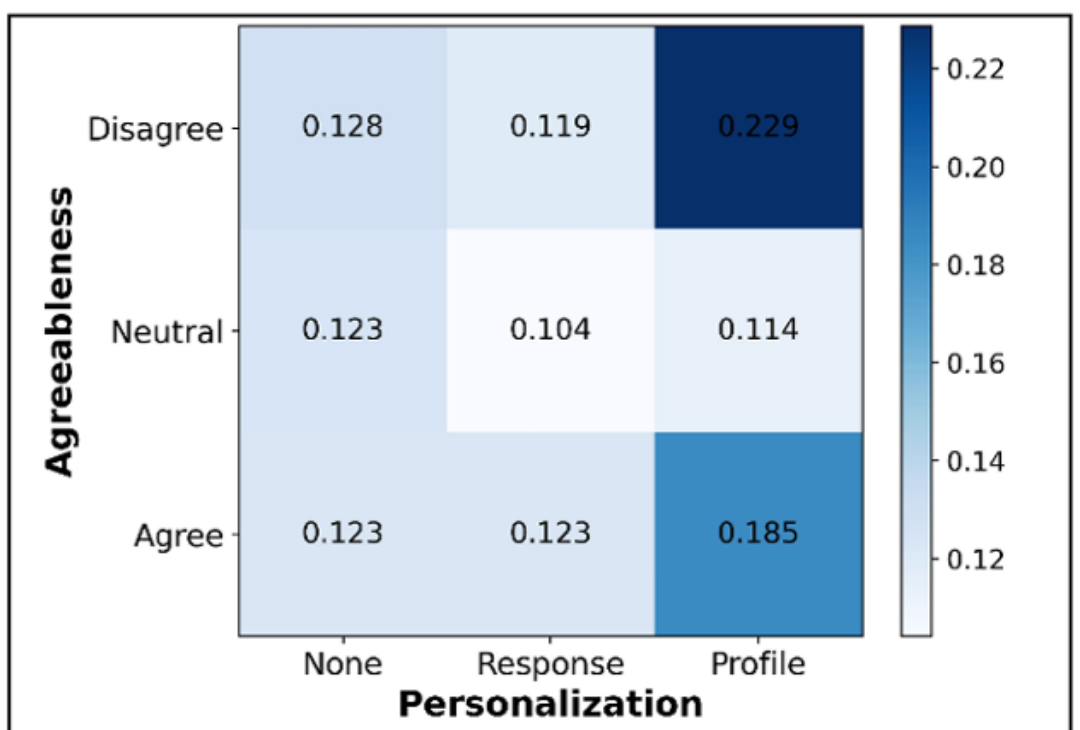

**Figure 5. Groups of primary contacts only.** Average difference in ratings using LLMs and ratings not using an LLM.

We also investigated model correlations as a form of robustness check as we expect outcomes to be correlated across models. When computing model-wise correlations for each setting shown in Figure 4, we obtain Table 1, which shows that alignment on an aggregate level is high. Interestingly, correlation is weakest between the European and the US model. There could be a number of reasons ranging from data filtering onto training procedures and model prompt following capabilities, e.g., weaker older models (like Ministral) might not be able to follow prompts as well as newer models like Qwen 3.5 and Gemma-4. (Correlations for settings in Figure 4 are similar)

| **Model** | **Mistral (Europe)** | **Qwen3.5 (Chinese)** | **Gemma-4 (US)** |
|---|---|---|---|
| Ministral (Europe) | 1 | 0.96 | 0.65 |
| Qwen3.5 (Chinese) | -(symmetric) | 1 | 0.82 |
| Gemma-4 (US) | -(symmetric) | -(symmetric) | 1 |

**Table 1. Correlation of models for differences as shown in Figure 3**

When looking at individual questions in Table 2, correlations are much lower. This is not unexpected as aggregates are in general less varying than individual samples (law of large numbers). Correlations of final responses of models are between 0.3 to 0.4, which could be categorized as being on the higher end of medium for psychological studies, where correlations are often below 0.2 (Lovakov & Agadullina, 2021).

| **Model** | **Ministral (Europe)** | **Qwen3.5 (Chinese)** | **Gemma-4 (US)** |
|---|---|---|---|
| Ministral (Europe) | 1 | 0.38 | 0.3 |
| Qwen3.5 (Chinese) | -(symmetric) | 1 | 0.38 |
| Gemma-4 (US) | -(symmetric) | -(symmetric) | 1 |

**Table 2. Correlation of models for differences computed for settings as in Figure 3 but computed on differences using individual statements**

To summarize, we find that personalization is the dominant driver of opinion formation, but only in conjunction with a non-neutral behavioral stance. Personalization alone, in the absence of an agreeable or disagreeable instruction, does not produce a meaningful effect. Once a non-neutral stance is present, however, providing the model with a user profile exerts a substantially stronger influence on group ratings than generic agreeableness instructions or the inclusion of a user's prior response (with differences significant at $p < 0.001$ in both agreeable and disagreeable conditions). In other words, agreeableness and personalization interact, but they do so asymmetrically. Profile-based personalization amplifies disagreeable instructions more strongly than agreeable ones. This asymmetry may be tied to the fact that model training already instills a baseline disposition toward agreement, making it less likely that this tendency is countered. Beyond these main effects, the composition of the group can also be seen as an important moderator. In homogeneous groups, where opinion reinforcement dynamics tend to push views toward more extreme positions, even a neutrally instructed LLM produces a substantially larger moderating effect than in heterogeneous groups ($p < 0.01$), underscoring that identical model behavior can yield qualitatively different opinion dynamics depending on the structure of the social context in which the model is embedded.

## Hypothesis Generation

Following a simulation paradigm where we substitute humans with LLMs, we can generate a number of observations and formulate a range of hypotheses about the influence of genAI on collective opinion dynamics. We do so along two dimensions; (1) ***AI-related features*** and how they are involved in collective opinion making, and (2) ***social configurations*** in which AI is having an effect.

### *AI-related Features*

The first set of hypotheses refer to features of AI. Thereby, we refer to design decisions that can be taken, for instance, to make an LLM more personal or credible. Overall, and in relation to AI-related features, our findings indicate that a user’s profile information matters much more than simply knowing the person's current response, particularly when the AI is instructed to agree or disagree. Following this, we propose four hypotheses.

### Personalization-based Influence

Our findings indicate that personalization can increase the influence of LLMs on collective opinion dynamics. In particular, providing the model with information about the user's profile leads to considerably larger differences from human-only discussions than conditions in which no such information is available. This effect is especially prevalent when the LLM adopts a non-neutral stance, that is, when it agrees or disagrees.

One possible explanation is that user profile-based information allows the LLM to generate arguments that are more closely aligned with characteristics of the individual and are therefore more persuasive. This observation is important, as it shows that the increasing availability of user information may allow LLMs to exert influence in ways that are less generic and more closely adapted to individual users (Bozdağ, 2025). We therefore propose that the influence of an LLM on individual and collective opinion formation increases with the degree to which its responses are personalized to user-specific information.

**$H_1$: *When LLM responses are personalized to a user's profile, they will exert greater influence on opinion making than generic responses.***

### Personalization and Non-neutral Stance

At the same time, our findings suggest that personalization does not exert the same influence under all conditions. When a user is provided with a personalized, yet neutral LLM, we observe comparatively limited changes. When the model agrees or disagrees, personalization has a larger effect. In short, personalization alone may not be sufficient to influence opinion dynamics, but its influence may depend on the stance that is propagated by the model.

From a broader perspective, this suggests that personalization can be understood as an amplifier. For instance, user information may provide an LLM with additional means to formulate and frame a given position in a way that resonates more with the user. Consequently, the influence of personalization should be considered together with the stance expressed by the LLM rather than as an isolated feature. We therefore propose that personalization facilitates the influence of genAI on opinion dynamics when an it adopts a non-neutral stance.

**$H_2$: *The influence of personalization will be stronger when the LLM adopts a non-neutral stance than when it remains neutral.***

### Disagreement and Personalization

Our findings further point to an asymmetry between 'agreeable' and 'disagreeable' model behavior. Across several conditions we observe that when an LLM disagrees, it leads to larger changes as compared than when it agrees.

This may be due to different reasons. First, users who already strongly agree or disagree with a statement have limited means or interest to move further in the same direction; an opposing position, however, can move them away from an existing extreme. Second, current LLMs may already exhibit a tendency toward agreement, such that explicit disagreement represents a more significant intervention. More generally, personalized disagreement may be particularly influential because it combines information about the user with a position that challenges the user's existing view. Such a combination may make opposing arguments more difficult to dismiss as irrelevant or unrelated to the user. We therefore propose that profile-personalized disagreement produces stronger changes in opinion than profile-personalized agreement.

**H3: *Disagreement based on a user profile will produce greater opinion change than profile-personalized agreement.***

### Depth of Personalization-related Information

Our findings suggest that personalization can be influential to different extents. Providing the LLM with the user's previous response alone seems to produce comparatively limited changes, while personalizing based on the user's characteristics (e.g., religious belief) results in stronger effects when the LLM takes on an agreeable or disagreeable stance.

This indicates that the influence of personalization may depend on the amount and type of information that is available to the LLM. Knowledge of a prior response informs the model about what a user is currently concerned with, whereas profile information provides a broader context about the characteristics on which an argument can be based or adapted. From this point of view, richer user information may increase the LLM's capacity to generate responses that appear more relevant and thus, convincing. We therefore propose that personalization based on larger amount of information exerts a stronger influence on opinion dynamics than personalization based solely on a user's previously stated opinion.

**$H_4$: *Personalization based on a broader user profile will produce greater opinion change than personalization based only on the user's previous stated opinion.***

## *Social Configurations*

The second set of hypotheses refers to the social context in which AI is used. To this end, we contrast heterogeneous networks (i.e., people encounter different perspectives) with homogeneous networks (i.e., consisting mainly of similar contacts). Overall, we find that the influence of AI on collective opinion dynamics depends on the extent to which social group provides diversity. On these grounds, we develop three hypotheses.

### Social Homogeneity

Our findings indicate that the social structure in which an LLM is used has a substantial influence on collective opinion dynamics. While overall, similar patterns emerge for both homogeneous and heterogeneous groups (i.e., neutral LLMs have the least influence, and profile-personalized LLMs which agree or disagree have the highest effect), the magnitude of these effects differs across the two social structures.

Most notably, neutral LLMs exert more influence in homogeneous groups than in heterogeneous groups. One possible explanation is that homogeneous groups reinforce existing opinions through repeated interaction with similar individuals. In such settings, even a neutral LLM introduces a perspective that differs from the prevailing group dynamics and thereby, potentially shifts collective opinions more considerably. In heterogeneous groups, by contrast, a neutral LLM has less influence. This may be because individuals are already exposed to more diverse perspectives through their social contacts. This suggests that the influence of an LLM does not depend solely on its features (see $H_1$-$H_4$), but also on the social context in which it is used. We therefore propose that LLMs exert stronger effects on collective opinion formation in homogeneous than in heterogeneous social groups.

**$H_5$: *LLM use will have greater influence on collective opinion making in homogeneous social networks than in heterogeneous networks.***

### Alternative Viewpoints

Another implication that follows is that the influence of an LLM may depend on the extent to which alternative viewpoints are available in a given social group. In our simulation, heterogeneous groups contain primary, secondary, and tertiary contacts and therefore expose individuals to people with more diverse profiles and potentially diverging perspectives; homogeneous groups, in contrast, primarily entail similar peers.

Hence, and in complement to $H_5$, we can also infer that an AI-generated perspective becomes more consequential when similar perspectives are not present in a given social group. In this sense, the diversity of human and AI-generated viewpoints is dependent on each other: when humans are already provided with exposure to different positions, an additional LLM perspective contributes comparatively little; when such an exposure is limited, however, the influence of an LLM becomes stronger. Importantly, our simulation does not focus on viewpoint diversity independent of network structure. Hence, this a speculative assumption that needs to be tested. Nevertheless, the reported differences between homogeneous and heterogeneous groups give us reasons to assume that the influence of an LLM potentially increases when the diversity of viewpoints through human interactions decreases.

**$H_6$: *An LLM will have more influence when the diversity of viewpoints available through human interaction decreases.***

**Disruption of Echo Chamber**

Furthermore, our findings indicate that LLMs may alter not only individual opinions but also the reinforcing dynamics in social groups through which collective opinions develop. In homogeneous social groups, interactions among similar individuals tend to reinforce existing views, leading to positions that drift from the neutral point. Introducing a neutrally instructed LLM weakens this tendency by moving group ratings toward more moderate positions.

In this respect, the LLM can function as an additional source of perspectives that are not generated by the immediate social group itself. Certainly, this does not imply that LLMs necessarily reduce polarization or that their outputs always represent viewpoints external to a group. Rather, the simulation suggests that an LLM can interrupt processes of within-group reinforcement when its responses differ from the trajectory in which the group tends to develop at given time. This is important because the influence of genAI may therefore arise not only from persuading individual users, but also from altering the interaction dynamics through which groups reproduce collective opinions. We therefore propose that, in socially homogeneous groups, the introduction of LLM-generated perspectives that differ from the group's prevailing trajectory weakens within-group opinion reinforcement.

**$H_7$: *In socially homogeneous groups, introducing an LLM that provides viewpoints outside the group's prevailing position will weaken within-group opinion reinforcement.***

# Implications

In this paper, we studied the influence of genAI on collective opinion formation. We explored two specific dimensions of LLMs, that have been associated with potentially high influence on user behavior and social outcomes: agreeableness and personalization. Previous research has shown that these dimensions are not only strategically implemented in LLMs, but they also affect how outcomes are perceived and interpreted (Jain et al., 2026; Kelley & Riedl, 2026). Importantly, these findings stress that the interaction behaviors of LLMs can even influence the extent to which users make judgements that go against social norms and moral values (Cheng et al., 2026).

In our simulation, we implemented these dimensions to different degrees and in different social settings and across a high number of interaction rounds. This allowed us to see whether agreeableness and personalization play a role in collective opinion making. Essentially, we find that LLMs can be considered to act as mediators that have the capability to potentially counteract forming of extreme, potentially threatening perspectives, if instructed to be neutral or disagreeing. Asking LLMs to be neutral has a profound impact that is not much different from asking to be disagreeing (except if profile information is used). Thus, as a practical implication, regulators might require LLM providers to add an instruction like "be neutral" to the system prompt to counteract extremist views. Under the assumption that only a few number of LLM providers exist, such a requirement can be much more easily controlled as global content on the internet and groups in social media. At the very same time, it also can lead to a reduction of diverse views.

Our findings are among the first ones to systematically understand opinion formation under the influence of LLMs. The hypotheses stated above can guide future research, particularly experimental one. Furthermore, future work could explore more nuanced network topologies that better reflect the spectrum of real-world social structures, moving beyond the polar cases of fully heterogeneous and homogeneous networks toward mixed connectivity patterns, which are more characteristic of contemporary societies.

Also, it can be studied what happens on much more aggregative levels, that is, how LLM-induced opinion formation can lead to social norms and conventions at much larger collective scales (e.g., nations or continents) (Grisold et al., 2025).

Finally, future work could also examine how varying adoption rates—including heterogeneous and context-dependent engagement patterns—shape the magnitude and distribution of LLM influence on opinion formation. Also, one can improve the fidelity with which LLMs simulate human respondents, as more behaviorally realistic agents would strengthen the validity of simulation-based approaches to studying human-AI interaction (Gao et al., 2025). Finally, expanding the scope and diversity of topics, particularly to include domains where controversy is more culturally situated or emotionally charged, would allow for a more comprehensive and representative assessment of how LLM-mediated interaction shapes collective opinion formation.

# Limitations

Our study is subject to limitations. Most importantly, we use LLMs to simulate human participants. While prior research indicates that LLM-based agents can reproduce aspects of human behavior (Park et al., 2023), such simulations also have weaknesses. For instance, they introduce systematic distortions, flatten out the heterogeneity of social identities, and can differ from the way actual humans would respond to social influence. Our findings should therefore not be taken as direct evidence of how human collectives will respond to LLMs, but rather as patterns that emerge under the assumptions that we implemented in our simulation. They require empirical testing with human participants, and our hypotheses can serve as starting points. Furthermore, opinion formation is intentionally conceptualized in a simplified manner through a five-point Likert-scale, as well as repeated interactions and a fixed memory of the four most recent encounters.

Importantly, the findings represent genAI's influence in terms of magnitude rather than the direction, that is, whether it shifts opinions toward agreement, disagreement, moderation, or extremity. Finally, a simulation simplifies complex phenomena by considering a limited number of conditions: personalization is represented through either user-profile information or a prior response, and social structure is modeled through homogeneous and heterogeneous networks. Certainly, real-world opinion formation involves more nuanced identities, relationships, network structures, histories, and situational contexts. The hypotheses should therefore be understood as theoretically informed propositions generated from a simulation rather than as established claims about real-world collective opinion formation.